\documentclass[a4paper]{article}

\usepackage{sty/INTERSPEECH2021}
\usepackage{adjustbox}
\usepackage{xcolor}
\usepackage[hidelinks]{hyperref}
\usepackage{caption}
\usepackage{subcaption}
\usepackage{cite}
\usepackage{stackengine}
\usepackage{multirow}
\usepackage{makecell}
\usepackage{arydshln}

\usepackage{cancel}

\title{On TasNet for Low-Latency Single-Speaker Speech Enhancement}

\name{Morten Kolbæk$^{1}$, Zheng-Hua Tan$^{1}$, Søren Holdt Jensen$^{1}$, Jesper Jensen$^{1,2}$}

\address{
	$^1$Department of Electronic Systems, Aalborg University, Denmark\\
	$^2$Oticon A/S, Denmark}
\email{kolbek@hotmail.com, \{zt,shj,jje\}@es.aau.dk, jesj@demant.com}

\begin{document}
\newcommand\barbelow[1]{\stackunder[1.2pt]{$#1$}{\rule{.8ex}{.075ex}}}

\maketitle

\begin{abstract}
In recent years, speech processing algorithms have seen tremendous progress primarily due to the deep learning renaissance. This is especially true for speech separation where the time-domain audio separation network\,(TasNet) has led to significant improvements.
However, for the related task of single-speaker speech enhancement, which is of obvious importance, it is yet unknown, if the TasNet architecture is equally successful.
In this paper, we show that TasNet improves state-of-the-art also for speech enhancement, and that the largest gains are achieved for modulated noise sources such as speech.   
Furthermore, we show that TasNet learns an efficient inner-domain representation, where target and noise signal components are highly separable. This is especially true for noise in terms of interfering speech signals, which might explain why TasNet performs so well on the separation task.  
Additionally, we show that TasNet performs poorly for large frame hops and conjecture that aliasing might be the main cause of this performance drop. 
Finally, we show that TasNet consistently outperforms a state-of-the-art single-speaker speech enhancement system.

\end{abstract}


\section{Introduction}
Algorithms capable of extracting a desired signal from a mixture of signals are useful for a large range of applications. For example, for mobile communication devices or hearing aids, it is desirable
to extract or enhance the speech signal of interest in order to limit the impact of background noise. Similarly, in applications involving videoconferencing and machine transcription of audio recordings it is useful to be able to separate the recorded speech signals such that intelligibility can be improved or preserved for both humans and machine receivers.

In recent years, following the advent of the deep learning renaissance, tremendous progress has been made within the speech processing field, as a large range of fairly successful speech enhancement and separation algorithms have been proposed (see, e.g., \cite{wang_supervised_2018,kolbaek_single-microphone_2018} and references therein). However, despite the fact that overlapping and concurrent speech rarely occurs in natural conversations \cite{shriberg_observations_2001,hilton_perception_2016}, algorithms designed to separate multiple speech signals have received a lot of recent attention. On the other hand, less attention has been devoted to algorithms designed to solve the, perhaps, more 
often encountered problem of enhancing a single-speaker speech signal corrupted by noise. 

In particular, the time-domain audio separation network\,(TasNet) \cite{luo_tasnet:_2018-1,luo_conv-tasnet:_2019} has drawn a lot of attention \cite{heitkaemper_demystifying_2020,kadioglu_empirical_2020} especially after it was shown \cite{luo_conv-tasnet:_2019} that TasNet could separate speech signals with higher fidelity than what was possible using the ideal ratio magnitude mask -- this was previously considered an ambitious goal for speech enhancement and separation algorithms \cite{wang_ideal_2005,hummersone_ideal_2014,wang_training_2014}.
Indeed, TasNet does perform significantly better than previous techniques, when evaluated on a classical speaker-independent multi-speaker speech separation task  (e.g., \cite{hershey_deep_2016,kolbaek_multi-talker_2017-1,chen_deep_2017}). However, it  is still unclear (e.g., \cite{heitkaemper_demystifying_2020,kadioglu_empirical_2020}), why this particular architecture performs so well on this task. Recently, the TasNet architecture has formed the basis for other algorithms improving the speech separation performance even further (e.g., \cite{luo_dual-path_2020,nachmani_voice_2020}). %

Despite the obvious importance of the single-speaker in noise task, as motivated above, the performance of TasNet for such speech enhancement task is yet unknown.
It is therefore of significant importance to establish, if TasNet also outperforms recent state-of-the-art single-speaker speech enhancement architectures  \cite{pandey_new_2018,kolbaek_loss_2020}.
To do so, we perform simulation experiments to compare TasNet with a state-of-the-art speech enhancement system based on the uNET architecture \cite{kolbaek_loss_2020}. We also study the inner-domain representation learnt by TasNet and demonstrate that speech and noise signals are more separable with this representation than with a traditional Short-Time Fourier Transform (STFT) representation. Finally, we demonstrate that TasNet performance drops significantly for large input frame hops - we conjecture that aliasing could play a role in this performance decrease.

\section{Single-Channel Speech Enhancement and Separation}
\label{sec:sc_se_ss}
Single-channel speech separation aims to separate a mixture of signals into its constituent parts with access only to a single-microphone recording of the mixture. Let $\barbelow{x}_i \in \mathbb{R}^L$ be $L$ samples of the $i^{th}$ noise-free time-domain speech signal and let $\barbelow{v} \in \mathbb{R}^L$ be an additive 
%non-speech 
% JJ: confusing since in exps (and in general), noise can 
% be competing speech. Do we loose anything by skipping?
noise signal. The noisy mixture signal $\barbelow{y} \in \mathbb{R}^L$ is then defined as $\barbelow{y} = \sum_{i}^{I}\barbelow{x}_i + \barbelow{v}.$ 
The general objective in deep-learning based speech separation is to find estimates $\hat{\barbelow{x}}_i$ of $\barbelow{x}_i$ from $\barbelow{y}$ using a deep neural network\,(DNN), $f_{DNN}(\barbelow{y},\mathbf{\barbelow{\theta}})$, parameterized with $\barbelow{\theta}$. Choosing $I=1$, the separation task boils down to the special case of a speech enhancement task.

\begin{figure}
	% trim={<left> <lower> <right> <upper>}
	\centering
	\includegraphics[trim={0mm 0mm 0mm 0mm},clip,width=1.0\linewidth]{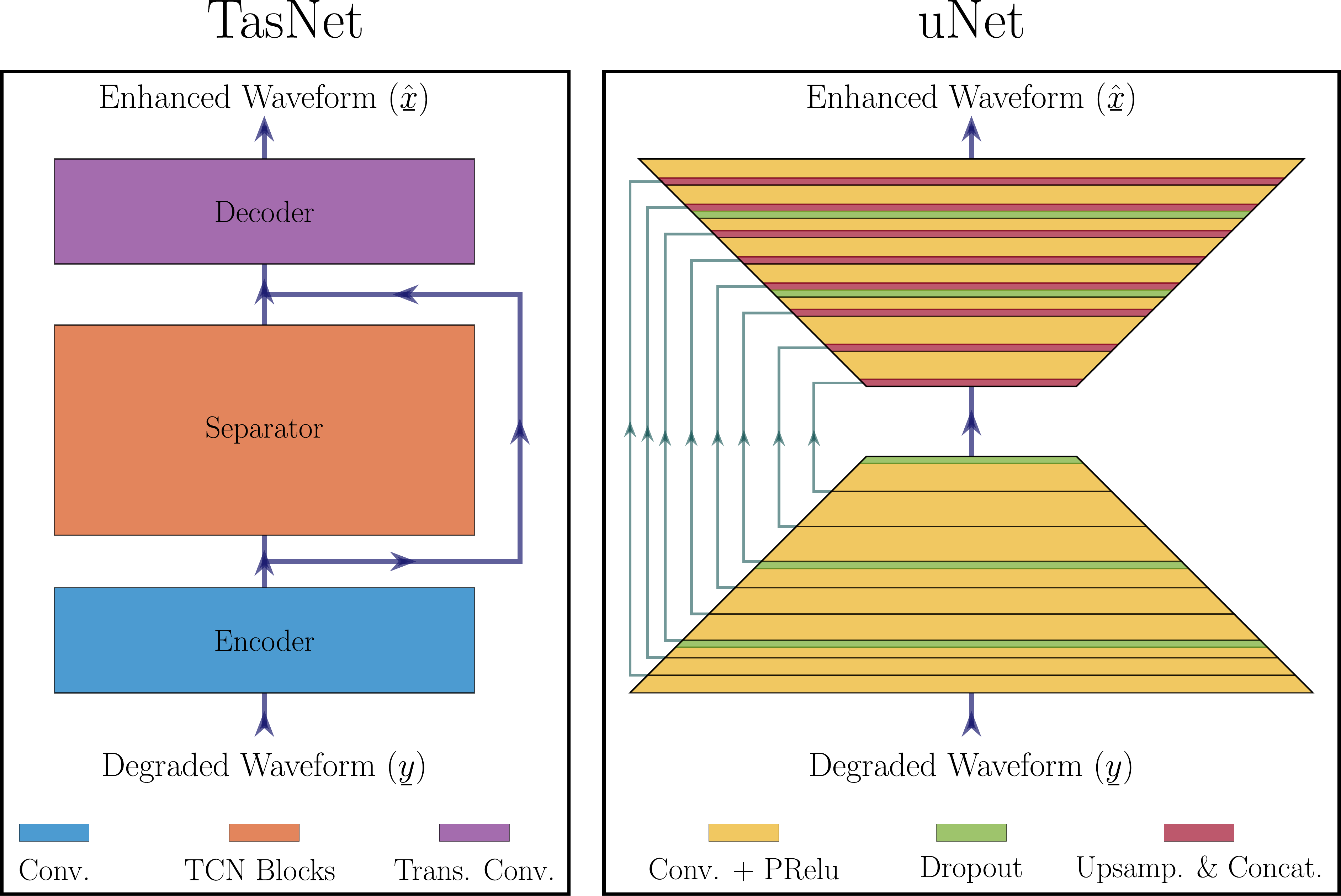}
	\caption{TasNet (left) and uNet (right) architectures.}
	\label{fig:sesys}
\end{figure}

In this study two deep-learning architectures are considered which are illustrated in Figure\;\ref{fig:sesys}; The TasNet architecture and the uNet architecture. Common to both architectures is that they are end-to-end time-domain models, which means that they both estimate $\hat{\barbelow{x}}_i$ directly from the noisy time-domain observation $\barbelow{y}$. 
The main difference is that TasNet is a speech separation model capable of separating multiple mixtures, i.e., $i=1,2,\dots, I$, whereas uNet is specifically designed for the speech enhancement task with only a single target speech signal, i.e., $I=1$.    
One goal of this study is to establish performance differences between $f_{TasNet}(\cdot,\cdot)$ and $f_{uNet}(\cdot,\cdot)$, respectively, for the speech enhancement (i.e., $I=1$) task.
\subsection{TasNet Architecture}\label{sec:tasnet_arch}
Figure\;\ref{fig:sesys} (left) shows the top-level TasNet architecture. The "Encoder" is a fixed (learnt) linear map, that maps successive noisy time-domain input frames to an inner-domain. Based on this inner-domain representation, the "Separator" estimates scalar weight values, which are applied point-wise to the inner-domain representation. Finally, the "Decoder", which is another fixed (learnt) linear map, transforms the weighted inner-domain representation to an enhanced time-domain output waveform $\hat{\barbelow{x}}$.

The TasNet architecture shows a strong resemblance to a traditional transform-based enhancement system. In particular, in traditional speech enhancement algorithms, the inner-domain would often be the STFT domain, i.e., the "Encoder" would be a fixed, linear transform, namely the Discrete Fourier Transform, applied to successive frames of the input waveform.
Denoising is achieved by point-wise multiplication of the inner-domain representation (the STFT coefficients) with scalar weights, estimated via a parametric statistical model (e.g., \cite{hendriks_dft-domain_2013}) or a DNN (e.g., \cite{kolbaek_single-microphone_2018,wang_supervised_2018}). Finally, the enhanced signal would be constructed via another fixed linear transform, namely the inverse STFT.

In TasNet, on the other hand, the  "Encoder" and "Decoder" are learnt in an end-to-end fashion, jointly with the separator.
The encoder and decoder are based on one-dimensional convolutions ("Conv" and "Trans. Conv" in Fig.\,\ref{fig:sesys}) and the separator network is based on a chain of multiple blocks of temporal convolutional networks\,("TCN") with increasing dilation \cite{luo_conv-tasnet:_2019}. 
%
%
%JJ: Why Trans.Conv? These are not transpose networks?
%MK: Then transpose in Trans. Conv simply means that we apply upsamling before applying the convolution. 
%

%JJ: is my ref above ok? or did you modify the original so much that it is now yours?
%MK: Yes, this ref is still valid. 

The TasNet implementation used in this work (adapted from \cite{wu_funcwjconv-tasnet_2021}) has three main configurations: 1) one with non-causal convolutions and with global layer normalization\,(gLN), 2) one with non-causal convolutions and cumulative layer normalization\,(cLN), and 3) one with causal convolutions and cLN. The first configuration is non-causal as the entire signal is used for gLN. The second configuration introduces a fixed system latency, because the non-causal convolutions require a signal look-ahead, which is dependent on kernel lengths.
%JJ is description of 2) above ok? can we say how much delay in our impl.?
%MK Yes, Yes, 
%
The latency of the third configuration is determined by the frame hop, which for most experiments is 1 ms.      
Adopting the notation introduced in \cite{luo_conv-tasnet:_2019}, we use the following configuration of TasNet: L16, K8, N512, X8, R3, B128, H512, P3, except where otherwise explicitly stated.
%JJ: I tried to explain notation above - is this correct? .
%MK: Yes, I think this is good.
With this configuration the TasNet model has 3.5 million parameters and a receptive field of 15,310 samples. 
For implementation and further details, we refer to \cite{luo_conv-tasnet:_2019,wu_funcwjconv-tasnet_2021}.

\subsection{uNET Architecture}
Figure\;\ref{fig:sesys} (right) shows the top-level uNet architecture.
uNet follows an autoencoder architecture with multiple strided convolutions and corresponding skip-connections. The uNet architecture was initially proposed for image processing \cite{ronneberger_u-net:_2015}, but the architecture was later shown to be successful in the speech domain and achieved state-of-the-art performance on time-domain speech enhancement (e.g., \cite{pandey_new_2018,park_fully_2017}).    
In (inChannel, outChannel, stride) format, uNet has one (1,48,1), two (48,48,2), one (48,96,2), two (96,96,2), one (96,180,2), two (180,180,2), two (180,180,1), one (180,96,1), two (96,96,1), one (96,48,1), two (48,48,1), and one (48,1,1) convolutional layers with a filter size of 11 samples. In this configuration uNet has 3.5 million parameters, which is similar to the chosen TasNet configuration. The receptive field is 2,561 samples.
Finally, all layers apply a parameterized ReLU activation function and dropout after every third layers in the encoder. For more details, we refer to \cite{kolbaek_loss_2020}.
%For further details please consult \cite{kolbaek_loss_2020}.

\section{Experimental Design}\label{sec:expdes}
In our simulation experiments, we mainly train and evaluate TasNet as an enhancement system, i.e., $I=1$, cf.\ Sec.\ \ref{sec:sc_se_ss}. In this case, TasNet is trained and evaluated using speech signals contaminated by additive noise. Network performance is evaluated in terms of STOI \cite{taal_algorithm_2011}, PESQ \cite{rix_perceptual_2001}, and Scale-Invariant SDR \cite{roux_sdr_2018}.  In some experiments, we also train and evaluate TasNet as a 2-speaker speech separation system ($I=2$) using  a speech  separation dataset. We do this to validate our TasNet implementation by comparing to results reported in literature. In this situation, the true source signals are used to determine the correct permutation of the TasNet output signals before performance is quantified in terms of mean STOI, PESQ, and SI-SDR.
%JJ: why not ESTOI?
%MK: Great question. 

\subsection{Speech Data}
The speech data used for all experiments is based on the WSJ0 speech corpus \cite{garofolo_csr-i_1993}. Specifically, the speech data used for training is %based on 
the \emph{si\_tr\_s} subset of WSJ0, which consists of 11,613 utterances, approximately equally divided among 44 male speakers and 47 female speakers. For validation the \emph{si\_tr\_s} subset is used, which consists of 1,163 utterances, divided among five male speakers and five female speakers, which are not present in the training set. For testing, the subsets \emph{si\_et\_05} and \emph{si\_dt\_05} are used, which consist of 1,857 utterances divided among ten males and six females. 
Furthermore, as we are primarily interested in speech active regions during training, we apply a voice activity detector that analyzes the clean waveform in 25 ms segments and removes the segments, where the signal energy is more than 40 dB below the energy of the segment with the maximum energy.
%in the waveform. 
Finally, all signals are concatenated or truncated to a length of 4 seconds and downsampled to 8 kHz. 
Datasets used for training contain 20,000 signals, whereas 2,000 signals are used for validation (wsj0-2mix is designed with 5,000 \cite{hershey_deep_2016}) and 3,000 signals are used for testing.

\subsection{Speech Mixtures for TasNet 2-Speaker Separation}
We have adopted the wsj0-2mix dataset \cite{hershey_deep_2016} as this is an often used dataset based on WSJ0 used for two-speaker speech separation. However, since the distribution of mixtures in wsj0-2mix is given as 17.7\% (530 mixtures) female-female mixtures, 28.9\% (867 mixtures) male-male mixtures, and 53.4\% (1603 mixtures) female-male mixtures it is not an accurate representation for the distribution of mixtures one would on average meet in real-life, which is $2 \times 25\%$ same-gender mixtures and 50\;\% opposite-gender mixtures. To study the performance on a balanced dataset, we introduce the wsj0-2bal dataset (denoted $2bal$ in Table\;\ref{tab:res1}), which has 50\;\% same-gender mixtures (25\;\% for males and 25\;\% for females) and 50\;\% opposite-gender mixtures to more accurately reflect a real-life scenario. For wsj0-2mix and wsj0-2bal, no noise is added to be comparable to most existing literature (see, e.g., \cite{kolbaek_joint_2017} for noisy wsj0-2mix separation). For wsj0-2mix and wsj0-2bal, the two speech signals are mixed at a random power ratio, selected uniformly from $[-2.5 , 2.5]$ dB.

\subsection{Noisy Speech for TasNet and uNET Enhancement}
The noisy speech signals are constructed by adding a noise-free utterance $\barbelow{x}$ with an equal length and randomly selected noise sequence $\barbelow{v}$. 
The noise sequence is selected from one of the following noise types:
stationary speech shaped noise\;($ssn$), competing female speaker\;($wsjf$), or the $mix$ noise, which is a concatenation of ssn, non-stationary 6-speaker babble and street, cafeteria, bus, and pedestrian noise signals from the CHiME3 dataset. For further information about the noise signals we refer to \cite{garofolo_timit_1993,barker_third_2015,kolbaek_speech_2017}. 
Note that when $wsjf$ is used, the corresponding clean speech data consists of male speakers only. This is to avoid the label permutation problem otherwise inhibiting training of uNet, 
see \cite{kolbaek_multi-talker_2017-1} for details.
For training the single-speaker speech enhancement systems, the SNR is chosen uniformly at random from $[-10 , 10 ]$ dB. For testing, an SNR of 0 dB is used. %The energy of the speech signals are determined using ITU P.56 \cite{itu_rec._2011}.    
%JJ: does it matter? or is it the SNR?

\subsection{Network Training}
TasNet and uNet are trained using the ADAM optimizer \cite{kingma_adam:_2015} with a learning rate schedule that reduces the learning rate with a factor of two, if the validation loss has not decreased for two epochs. TasNet uses a learning rate of $0.001$ and uNet uses a learning rate of $0.0001$. A batch size of eight is used for both TasNet and uNet, and training is stopped, if the validation loss has not decreased for five epochs or a maximum of 100 epochs has elapsed for TasNet and 200 epochs has elapsed for uNet. 
\section{Experimental Results}
\begin{table*}[th]
	\caption{STOI, PESQ, and SI-SDR scores for TasNet and uNet trained and tested with various datasets. }
	\label{tab:res1}
	\begin{adjustbox}{width=1.0\textwidth}
	\renewcommand{\arraystretch}{1.02}
	\centering
		\begin{tabular}{ l l c c c c c c c c c c c}
		\toprule
		& & \multirow{ 2}{*}{\makecell{\#Target \\ speakers}} & \multirow{ 2}{*}{\makecell{Receptive \\ field (s)}} & & & \multicolumn{2}{c}{\textbf{STOI}} & \multicolumn{2}{c}{\textbf{PESQ}} & \multicolumn{2}{c}{\textbf{SI-SDR}} \\ \cline{7-12}
		Test set & Model & & & Norm & Causal & Noisy & Processed & Noisy & Processed & Noisy & Processed \\ \midrule 		
        2mix & TasNet &2& 1.53 & gLN & N & 0.74 & 0.94 & 1.68 & 2.96 & 0.0 & 14.1 \\ 
        2mix & TasNet &2& 1.53 & cLN & N & 0.74 & 0.93 & 1.68 & 2.93 & 0.0 & 13.0 \\ 
        2mix & TasNet &2& 1.53 & cLN & Y & 0.74 & 0.90 & 1.68 & 2.52 & 0.0 & 10.2 \\   \hdashline
        2bal & TasNet &2& 1.53 & gLN & N & 0.70 & 0.94 & 1.50 & 2.84 & 0.0 & 13.4 \\ 
        2bal & TasNet &2& 1.53 & cLN & N & 0.70 & 0.91 & 1.50 & 2.69 & 0.0 & 11.9 \\ 
        2bal & TasNet &2& 1.53 & cLN & Y & 0.70 & 0.89 & 1.50 & 2.29 & 0.0 & 9.4 \\   \toprule
        mix & TasNet &1& 1.53 & gLN & N & 0.76 & 0.93 & 1.56 & 2.64 & 0.0 & 12.2 \\ 
        mix & TasNet &1& 1.53 & cLN & N & 0.76 & 0.93 & 1.56 & 2.63 & 0.0 & 12.3 \\ 
        mix & TasNet &1& 1.53 & cLN & Y & 0.76 & 0.89 & 1.56 & 2.17 & 0.0 & 9.7 \\  
        mix & uNet &1& 0.32 & n/a  & N & 0.76 & 0.91 & 1.56 & 2.41 & 0.0 & 11.1 \\  \hdashline
        ssn & TasNet &1& 1.53 & gLN & N & 0.73 & 0.93 & 1.42 & 2.56 & 0.0 & 11.4 \\ 
        ssn & TasNet &1& 1.53 & cLN & N & 0.73 & 0.93 & 1.42 & 2.57 & 0.0 & 11.4 \\ 
        ssn & TasNet &1& 1.53 & cLN & Y & 0.73 & 0.89 & 1.42 & 2.11 & 0.0 & 9.1 \\  
        ssn & uNet &1& 0.32 & n/a  & N & 0.73 & 0.91 & 1.42 & 2.34 & 0.0 & 10.2 \\  \hdashline
        wsjf & TasNet &1& 1.53 & gLN & N & 0.71 & 0.96 & 1.53 & 3.19 & 0.1 & 15.2 \\ 
        wsjf & TasNet &1& 1.53 & cLN & N & 0.71 & 0.97 & 1.53 & 3.33 & 0.1 & 16.0 \\ 
        wsjf & TasNet &1& 1.53 & cLN & Y & 0.71 & 0.95 & 1.53 & 2.74 & 0.1 & 12.4 \\   
        wsjf & uNet   &1& 0.32 & n/a & N & 0.71 & 0.95 & 1.53 & 2.81 & 0.1 & 13.4 \\   \hdashline
		\end{tabular}
	\end{adjustbox}
\end{table*}
\begin{table*}[th]
	\caption{STOI, PESQ, and SI-SDR scores for TasNet with varying window and hop sizes.}
	\label{tab:res3}
	\begin{adjustbox}{width=1.0\textwidth}
		\renewcommand{\arraystretch}{1.02}
		\centering
		\begin{tabular}{ l l c c c c c c c c c c c}
			\toprule
			& & \multirow{ 2}{*}{\makecell{\#Target \\ speakers}}  & \multirow{ 2}{*}{\makecell{Window/ \\hop [ms]}} & & & \multicolumn{2}{c}{\textbf{STOI}} & \multicolumn{2}{c}{\textbf{PESQ}} & \multicolumn{2}{c}{\textbf{SI-SDR}} \\ \cline{7-12}
			Test set & Model & &  & Norm & Causal & Noisy & Processed & Noisy & Processed & Noisy & Processed \\ \midrule 
            2bal & TasNet &2& 2/1 & gLN & N & 0.70 & 0.93 & 1.50 & 2.66 & 0.0 & 12.4 \\ 
            2bal & TasNet &2& 64/32  & gLN & N & 0.70 & 0.79 & 1.50 & 1.71 & 0.0 & 4.8 \\ 
            2bal & TasNet &2& 64/1  & gLN & N & 0.70 & 0.93 & 1.50 & 2.78 & 0.0 & 12.5 \\ \toprule
            ssn & TasNet &1& 2/1 & gLN & N & 0.73 & 0.93 & 1.42 & 2.60 & 0.0 & 11.4 \\ 
            ssn & TasNet &1& 64/32  & gLN & N & 0.73 & 0.87 & 1.42 & 2.07 & 0.0 & 8.2 \\ 
            ssn & TasNet &1& 64/1  & gLN & N & 0.73 & 0.93 & 1.42 & 2.60 & 0.0 & 11.0 \\ \toprule
		\end{tabular}
	\end{adjustbox}
\end{table*}
\begin{figure}
	% trim={<left> <lower> <right> <upper>}
	\centering
	\includegraphics[trim={10mm 4mm 10mm 2mm},clip,width=1.0\linewidth]{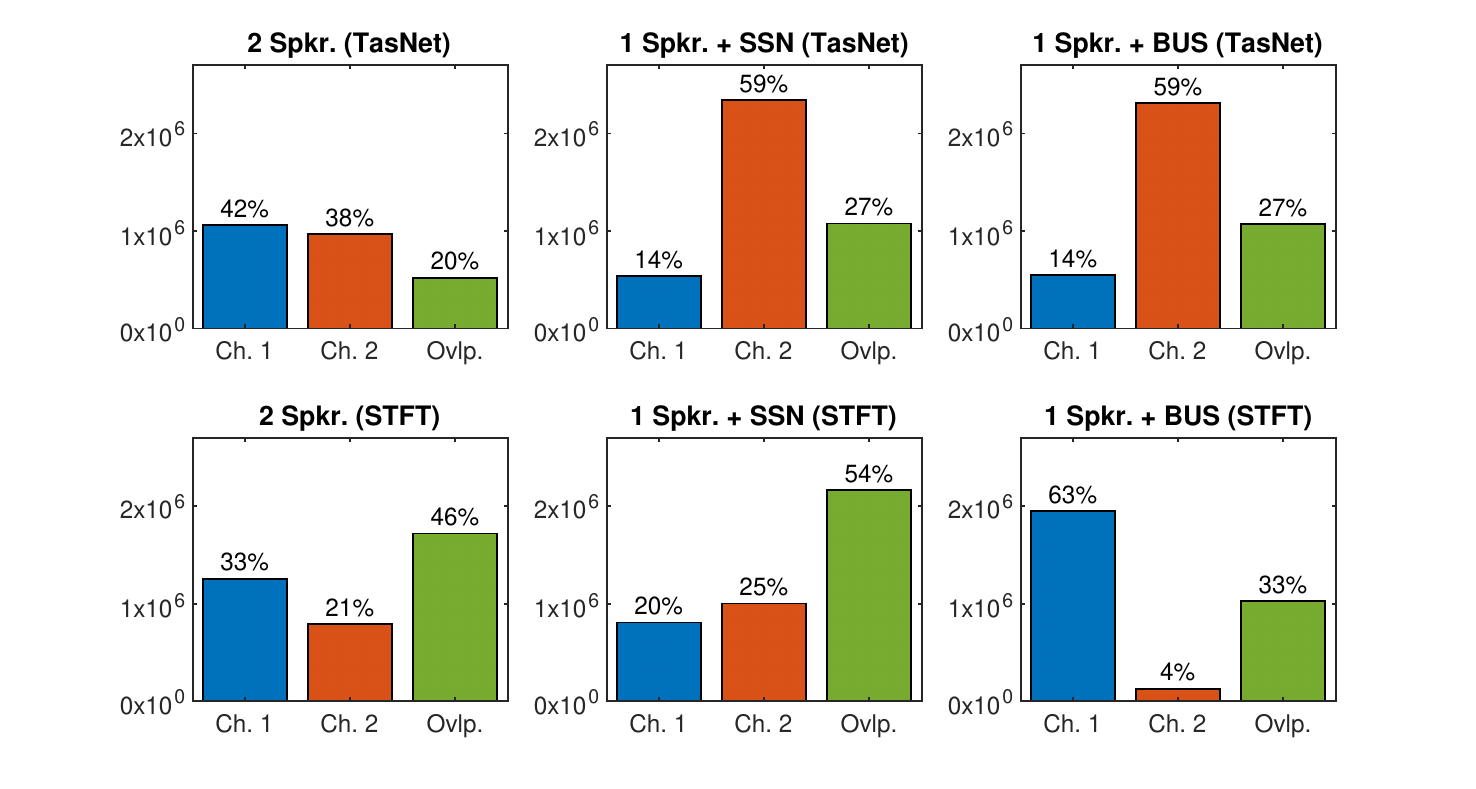}
	%\caption{Percentages indicating overlap between two signal in either STFT domain or TasNet domain. Generally, the TasNet domain achieves the largest separation.}
	\caption{Signal overlap in STFT domain and TasNet domain.}
	\label{fig:overlap}
\end{figure}
Table\;\ref{tab:res1} shows STOI, PESQ, and SI-SDR scores for signals processed by different TasNet systems. 
\subsection{TasNet 2-Speaker Separation Performance}\label{sec:res1}

First, in order to validate our implementation of TasNet, we focus on the 2-speaker separation task, see Table\;\ref{tab:res1} (rows with test sets $2mix$ and $2bal$). When TasNet is trained and tested on $2mix$ and with gLN normalization and non-causal convolutions, a STOI score of 0.94, PESQ of 2.96 and SI-SDR of 14.1 dB are achieved. This corresponds well with the reported SI-SDR score of a similar system in \cite{luo_conv-tasnet:_2019}, validating the trained TasNet model.  
Next, a TasNet separation network, trained and tested on $2bal$ and with gLN normalization and non-causal convolutions, achieves a STOI of 0.94, PESQ of 2.84 and SI-SDR of 13.4 dB, which for PESQ and SI-SDR is considerably less than with $2mix$. We argue, that the results with $2bal$ better reflect the performance to be expected in a gender-balanced real-life situation, whereas evaluation results with the $2mix$ test set are biased, and generally over-optimistic, wrt. real-life performance. The reason for this is that opposite-gender mixtures, which are over-represented in $2mix$, are easier to separate than same-gender mixtures and that male-male mixtures in general are easier to separate than female-female mixtures \cite{wang_pitch-aware_2019,kolbaek_multi-talker_2017-1}. 
\subsection{TasNet Speech Enhancement Performance}
Next, we consider TasNet for single-talker enhancement (i.e., rows in Table\;\ref{tab:res1} with Test sets $mix$, $ssn$, and $wsjf$). Using cLN instead of gLN as normalization in TasNet generally results in a small performance drop, whereas also using causal convolutions -- and, hence, implementing an essentially causal system -- leads to a much larger drop. Also, the drop is largest for non-stationary noise sources, such as the competing female speaker in test set $wsjf$. This could suggest that access to future information is particularly advantageous, when noise sources are non-stationary (such as a competing speaker), whereas the advantage is smaller for stationary noise types such as $ssn$.

Table\;\ref{tab:res1} also shows that TasNet with cLN normalization and causal convolution in general performs slightly better in terms of STOI, PESQ, and SI-SDR compared to uNet, which also uses non-causal convolutions. Since the models have the same number of parameters, this suggests that TasNet is a more efficient architecture, potentially due to the larger receptive field, see Sec.\ \ref{sec:sc_se_ss}.

\subsection{TasNet Inner Domain Analysis}\label{sec:res2}
In order to explore reasons for TasNets superior performance, we study the signal representation in the TasNet inner domain (cf. Sec. \ref{sec:tasnet_arch}). Specifically, we compute the inner domain overlap, i.e., the number of inner domain coefficients, where both target and noise signals contribute significantly, relative to the total number of coefficients. A low inner domain overlap indicates that signal and noise inner domain representations are disjoint and, hence, that noise may be eliminated without harming the target speech signal.
Figure\;\ref{fig:overlap} shows examples of inner domain overlaps for a competing speaker, ssn, and bus noise for TasNet (top row) and for an STFT filterbank (bottom row).
From Figure\;\ref{fig:overlap} (bottom row) it is clear bus noise has little overlap with the target speech - this is because the bus noise is dominated by low frequency content. The competing speaker and ssn has significantly larger overlap with the target speech in the STFT domain.
From Figure\;\ref{fig:overlap} (top row), the inner domain overlap with TasNet is reduced compared to the STFT representation. In particular, the overlap is significantly reduced for a target speech signal contaminated by a competing speaker, but also for ssn. For bus, the overlap reduction is smaller, presumably because of the lowpass nature of the bus noise, which makes the STFT representation quite efficient in the first place.
We hypothesize that the ability of TasNets "Encoder" to represent input signals with small overlap in the inner-domain could be a key reason for TasNets success.

\subsection{TasNet - Impact of Window and Hop Sizes}\label{sec:res3}
Table\;\ref{tab:res3} shows STOI, PESQ, and SI-SDR scores for signals processed by TasNet systems configured with varying window- and hop sizes and tested with $2bal$ and $ssn$. 
Clearly, short windows (2ms) and short hops (1ms) lead to good performance, while performance drops for long windows (64ms) and hops (32ms). 
However, with a hop size of 1 ms and a 64ms window, the performance is regained.
This latter result might be expected as the 64/1 system is a generalization of the 2/1 system - however, the result clearly emphasizes that hop size and not window length is the critical hyperparameter.
These results could be explained by the fact that aliasing is introduced, when the hop size is larger than one sample. To see this, recall that the "Separator" in TasNet is implemented using convolutions, which operate independently on separate rows of the inner domain representation (a row being defined as a particular inner domain coefficient, as a function of time) \cite{luo_conv-tasnet:_2019}. These rows are effectively decimated versions of the time-domain input signal with a decimation factor equal to the hop size in samples. However, as no anti-aliasing filters are applied, these inner domain rows contain aliasing components \cite{gong_impact_2018}. Hence, a larger hop size might lead to larger aliasing. For $2bal$, the aliasing components would consist of high frequency harmonics of the competing speaker and of the target speaker, whereas for $ssn$, they consist of mirrored high-frequency ssn and harmonics of the target speaker. 
Table\;\ref{tab:res3} suggests that a large hop size is much more harmful for a competing speaker situation ($2bal$) than for the $ssn$ situation. We hypothesize that this is so, because the aliasing components introduced with $ssn$ are broadband and more disperse and hence more similar to the noise seen during system training.
The presented aliasing hypothesis is supported by recent works (e.g. \cite{luo_dual-path_2020,nachmani_voice_2020}) which show that TasNet can be significantly improved if the "Separator" is allowed to operate {\em across} rows in the inner-domain. Operating across rows could allow the "Separator" to correct for the aliasing as it gets access to the entire signal, and could, hence, explain why these new architectures outperform the original TasNet architecture.

\section{Conclusion}
In this paper we study aspects of TasNet for single-channel speech enhancement. We demonstrate that TasNet consistently outperforms a state-of-the-art uNET-based speech enhancement system of similar complexity. Hence, the excellent performance of TasNet on speech separation tasks is maintained for speech enhancement tasks.
We also show that TasNet learns an efficient inner domain signal representation, where speech and noise components are highly separable - this is particularly so, for competing speech signals, explaining the excellent speaker separation performance of TasNet. Finally, we show that hop size, but not window size, is crucial for good Tasnet performance. We conjecture that large hop sizes introduce aliasing, which leads to deriorated TasNet performance.

%\balance
\cleardoublepage
\bibliographystyle{bib/IEEEtran}
%\bibliography{bib/mybib}
\bibliography{bib/mybib_1,bib/mybib_2}

\end{document}